\documentclass{article}
\usepackage[utf8]{inputenc}
\usepackage{graphics}
\usepackage{graphicx}
\usepackage[margin=2cm]{geometry}
\usepackage{caption}
\usepackage{subcaption}
\usepackage{color}
\usepackage{amsfonts}

\newcommand{\nd}{\noindent}

\title{Probabilistic Models with Nonlocal Correlations: Numerical Evidence of $q$-Large Deviation Theory}
\author{\small{D. J. Zamora$^{1,2}$\thanks{E-mail: javierzamora055@gmail.com}, C. Tsallis$^{1,3,4}$\thanks{E-mail: tsallis@cbpf.br}}, \\
\small{$^1$ Centro Brasileiro de Pesquisas Fisicas and}\\
\small{ National Institute of Science and Technology for Complex Systems,}\\
\small{ Rua Dr. Xavier Sigaud 150, Rio de Janeiro, 22290-180, Brazil.}\\
\small{$^2$ Instituto de Fisica del Noroeste Argentino,}\\
\small{ Av. Independencia 1800, Tucuman, CP 4000, Argentina}\\
\small{$^3$ Santa Fe Institute, 1399 Hyde Park Road, Santa Fe, 87501, NM, United States}\\
\small{$^4$  Complexity Science Hub Vienna, Josefstadter Strasse 39, Vienna, 1080, Austria}}

\date{\today}

\begin{document}

\maketitle

\begin{abstract}
The correlated probabilistic model introduced and analytically discussed in Hanel et al (2009) is based on a self-dual transformation of the index $q$ which characterizes a current generalization of Boltzmann-Gibbs statistical mechanics, namely nonextensive statistical mechanics, %constructed upon nonadditive %entropic functionals.
%the self-dual transformation $\bar q \to \frac{7-5q}{5-3q}$, hence $ q \to \frac{7-5\bar q}{5-3\bar q}$,
and yields, in the $N\to\infty$ limit, a $Q$-Gaussian distribution for any chosen value of $Q \in [1,3)$. 
We show here that, by properly generalizing that self-dual transformation, it is possible to obtain an entire family of such probabilistic models, all of them 
%using the generalized self-dual transformation $\bar q \to \frac{(c+2)-(c+1)q}{(c+1)-cq}$, it is possible to introduce a family of such models, 
yielding $Q_c$-Gaussians ($Q_c \ge 1$) in the $N\to\infty$ limit.
%with $Q_c=\frac{(5-2c)Q_c+2c-3}{(3-2c)Q_c+2c-1}$. 
This family turns out to be isomorphic to the Hanel et al model through a specific monotonic transformation $Q_c(Q)$.
%$(Q_c,c) \to (Q,3/2)\;(c \ge 0)$. 
Then, by following along the lines of Tirnakli et al (2022), we numerically show that
this family of correlated probabilistic models provides further evidence towards a $q$-generalized Large Deviation Theory (LDT), consistently with the Legendre structure of thermodynamics. The present analysis deepens our understanding of complex systems (with global correlations among their elements), supporting the conjecture that generic  models whose attractors under summation of $N$ strongly-correlated random variables are $Q$-Gaussians, might always be concomitantly associated with  $q$-exponentials in the LDT sense.
\end{abstract}


\section{Introduction}

Boltzmann-Gibbs (BG) statistical mechanics provides several cornerstone relations, of which the Maxwellian distribution of velocities and the exponential distribution of energies are the most important ones \cite{Reif}.
This is reflected in the Central Limit Theorem (CLT) \cite{Billingsley,Billingsley95} which leads, when the number $N$ of involved random variables increases indefinitely, to convergence towards normal distributions, and to the Large Deviation Theory (LDT) \cite{Varadhan66,Varadhan84,Ellis85,Hollander08,Touchette09} which describes the speed at which Gaussians are approached as $N$ increases.

The Gaussian is the $N\rightarrow\infty$ attractor of the appropriately centered and scaled sum of $N$ independent (or weakly correlated) discrete or continuous random variables whose second moment is finite (CLT). The simplest probabilistic model that realizes these paradigmatic properties is a set of $N$ independent equal binary random variables (each of which, say, takes the values 0 and 1).

Consider then a binary stochastic system with $N$ random variables yielding $n$ times $0$, and $(N-n)$ times $1$. In the LDT, we are concerned with the probability $P_N(n/N > z)\in[0,1]$ of the random variable $n/N$ taking values greater than a fixed value $z\in\mathbb{R}$ for increasingly large values of $N$. Under the hypothesis of probabilistic independence, we expect $P_N$ to behave like an exponential function, i.e., $P_N(n/N > z)\sim e^{-r_1(z)N}$, where the rate function $r_1$ corresponds to a BG relative entropy per particle;  $r_1(z)N$ plays the role of the total thermodynamic entropy, which is extensive in agreement with the Legendre structure of classical thermodynamics, i.e., $r_1(z)N\propto N\;(N> > 1)$.

The standard LDT reflects the BG statistical mechanics, which describes the thermal equilibrium of short-range Hamiltonian systems with Maxwellian velocity distribution. Within this theory, exponential distributions emerge naturally. It is generally applicable to a large class of relevant systems satisfying the Central Limit Theorem (CLT), including dynamical systems whose maximal Lyapunov exponent is positive, which guarantees strong chaos and, therefore, mixing in phase space and ergodicity. This fact enlarges the applicability of Markovian processes. We remind that the CLT focuses on sums of the random variables. This is directly related to temporal averages, which are of essential interest for statistical physics given their connection with ensemble averages through the ergodic hypothesis.

In this paper, we are interested in what happens if these binary random variables are not independent (nor nearly so) and the correlations between them are strong enough. In principle, there is no reason to expect that the corresponding limiting distribution is a Gaussian. In fact, if the attractors are not Gaussian, we are not inside the BG-statistics framework. If we focus on stationary states of typical complex systems, both the CLT and the LDT must be generalized. In nonextensive statistical mechanics \cite{Tsallis88,Beck03,Hanel14,Tsallis09} we typically deal with a wide class of strongly correlated systems. The associated velocity distributions appear frequently to be Q-Gaussian ones with $Q>1$ \cite{Anteneodo98,Cirto14,Cirto18,Rodriguez19}, with $Q=1$ when the interactions are short-ranged. These facts are associated with the so-called $Q$-Central Limit Theorem ($Q$-CLT) which, when $N\rightarrow\infty$, leads to convergence to a $Q$-Gaussian distribution $f(z)=e_Q(-z^2)$ \cite{Umarov}. We remind that $e_Q(z)=[1+(1-Q)z]^\frac{1}{1-q}$, with $e_1(z)=e^z$. If the attractors are $Q$-Gaussians then we expect the LDT to yield a $q$-exponential function, with $q(Q)$ being a monotonic function of $Q$ such that $q(1)=1$.

We focus here on a scale-invariant stochastic process with strongly correlated exchangeable random variables, known to yield a long-tailed $Q$-Gaussian attractor in the space of distributions. This is a class of correlated processes that can be interpreted as mean-field models which might be relevant in artificial, social, and natural systems. For example, numerical indications for the distributions of velocities in quasistationary states of long-range Hamiltonians suggest q-Gaussians \cite{Anteneodo98,Pluchino07}. Experimental and observational evidence for q-Gaussians exist for the motion of biological cells \cite{Thurner03} defect turbulence \cite{Daniels04}, solar wind \cite{Burlaga05}, among others.

In \cite{Tirnakli2022}, it was shown that the corresponding
LDT probability distribution is given by $P(N,z)=P_0\,e_q^{-r_q(z).N}$ with $q=2-1/Q\in(1,5/3)$ for a scale-invariant stochastic process. The probabilistic model used, introduced in \cite{HanelThurnerTsallis2009}, is based on the self-dual transformation $\bar q \to \frac{7-5q}{5-3q}$, hence $ q \to \frac{7-5\bar q}{5-3\bar q}$. In that work, another transformation is suggested as an alternative, namely,  $\bar q \to \frac{5-3q}{3-q}$, hence $q \to \frac{5-3\bar q}{3-\bar q}$. We show here that, by using the generalized self-dual transformation \cite{Tsallis2017,GazeauTsallis2019}

\begin{equation}
\bar q= \frac{(c+2) – (c+1)\,q}{(c+1)-c\,q}= \frac{(\frac{c}{2}+1) – (\frac{c}{2}+\frac{1}{2})\,q}{(\frac{c}{2}+\frac{1}{2})-\frac{c}{2}\,q}  \,,
\label{generalqbar}
\end{equation}

\nd it is possible to introduce a family of such models. 

\section{Model and results}

In the present paper, we focus on a scale-invariant probabilistic family of models for exchangeable stochastic processes. The random variables are binary (Ising-like) with correlated elements say from $x\in \{0,1\}$. By exchangeable we mean that the N-point probabilities $p_N(x_1, x_2,...,x_N)$ are totally symmetric in their arguments for all $N$, and that $p_N$ can be obtained by marginalization of $p_{N+1}$. Particularly, the probability of a specific microstate $(x_1, x_2,...,x_N)$ does not depend on the order of binary events, but only on the number $n$ of events in the state $x=0$ and $(N-n)$ events in the state $x=1$. We denote this probability as $r_n^N$. 

The probabilistic model in \cite{HanelThurnerTsallis2009}, through the Laplace-de Finetti theorem for exchangeable random variables, yields

\begin{equation}
    r^N_n=\frac{B\left(\frac{3/2-Q/2}{Q-1}+n,\frac{3/2-Q/2}{Q-1}+N-n\right)}{B\left(\frac{3/2-Q/2}{Q-1},\frac{3/2-Q/2}{Q-1}\right)}\,,
    \label{rnN}
\end{equation}
where $B(x,y)$ is the Euler Beta function.

We start by generalizing the transformations $\bar q(q)$ in \cite{HanelThurnerTsallis2009} as described in Eq. (\ref{generalqbar}). If we take $c=3/2$ we obtain $\bar q=\frac{7-5q}{5-3q}$,  indicated in \cite{HanelThurnerTsallis2009}. If we take $c=1/2$ we obtain $\bar q=\frac{5-3q}{3-q}$, also indicated in \cite{HanelThurnerTsallis2009} as an alternative possibility to be associated with a probabilistic model including nontrivial correlations.

We see in transformation (\ref{generalqbar}) that $c$ plays the role of $3/2$ within \cite{HanelThurnerTsallis2009}. Consistently we heuristically propose, generalizing Eq. (\ref{rnN}), the Ansatz described here below.

A specific micro-state with $n$ values 0 and $(N-n)$ values
1 corresponds, for $Q=1$,  to $r^N_n=1/2^N$ ; and for long-tailed distributions:

\begin{equation}
    r^N_n=\frac{B\left(\frac{c-(c-1)Q_c}{Q_c-1}+n,\frac{c-(c-1)Q_c}{Q_c-1}+N-n\right)}{B\left(\frac{c-(c-1)Q_c}{Q_c-1},\frac{c-(c-1)Q_c}{Q_c-1}\right)},
    \label{beta}
\end{equation}
The case in \cite{HanelThurnerTsallis2009} clearly corresponds to $c=3/2$.
In the present model there are $ N!/[n!(N-n)!]$ equivalent such micro-states $(n=0,1,2,...,N)$. Consistently, we have

\begin{equation}
    \sum_{n=0}^N\frac{N!}{n!(N-n)!}r^N_n=1.
\end{equation}

Following \cite{HanelThurnerTsallis2009}, we define,

\begin{equation}
    u^N_n\equiv \frac{(n/N-1/2)}{\sqrt{(Q_c-1)(n/N)(1-n/N)}},
\end{equation}

\nd and also

\begin{equation}
    \tilde u^N_n\equiv \frac{u^N_n}{2\, max_{n=1,2,...,N-1}u^N_n}.
\end{equation}

We also define the discrete width

\begin{equation}
    du^N_n\equiv \frac{[(n/N)(1-N/n)]^{-3/2}}{4(N+1)\sqrt{Q_c-1}},
\end{equation}

\nd from which it follows the un-normalized distribution

\begin{equation}
    F^N_n=(du^N_n)^{-1}\frac{N!}{n!(N-n)!}r^N_n.
\end{equation}

After normalization we have

\begin{equation}
    \tilde F^N_n\equiv \frac{F^N_n}{\sum_{n=1}^{N-1}F^N_n}.
\end{equation}

\begin{figure}[h!]
    \centering
    \begin{subfigure}[b]{0.49\textwidth}
        \centering
        \includegraphics[width=\linewidth]{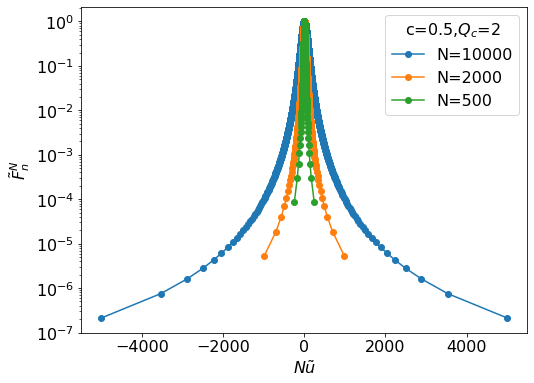}
        \caption{}
        \label{fig2c05Q25}
    \end{subfigure}
    \hfill
    \begin{subfigure}[b]{0.49\textwidth}
        \centering
        \includegraphics[width=\linewidth]{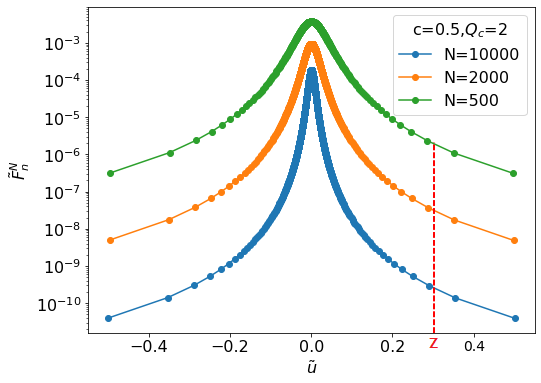}
        \caption{}
        \label{fig1c05Q25}
    \end{subfigure}
    \caption{ $F^N_n$ distributions are given for some representative values of $N$ for the case $(c=0.5,Q_c=2)$. The sums of the values of all points equals unity. Notice a relevant point, namely that the abscissa values of these points are not equidistant. (a) The distribution is represented as a function of $N\tilde u\in[-N/2,N/2]$. (b) The distribution is represented as a function of $\tilde u\in[-1/2,1/2]$. A red vertical line representing $z=0.3$ is plotted to illustrate that for smoothing the $P(N;Q,z)$ curve one has to interpolate between points.}
    \label{fig3}
\end{figure}

As an illustration, in Fig. \ref{fig2c05Q25} we have represented the data $(N\tilde u^N_n,\tilde F^N_n)$ for the case $(c=0.5,Q_c=2)$, whereas in Fig. \ref{fig1c05Q25} the same data have been plotted with respect to $\tilde u$ so that the distribution can be given in the region $[-1/2,1/2]$.

In what concerns LDT, we now focus on the probability $P(N;Q,z)\in [0,1]$, which is defined as the one whose values of $F^N_n$ correspond to $n/N > 1/2+z$. More precisely, it is the sum of all values whose $\tilde u > z$. We expect to numerically verify that

\begin{equation}
    P(N;Q,z)=P_0(Q,z)\,e_q^{-r_q(Q,z)N}.
    \label{qexp}
\end{equation}

Since the argument of the Beta functions in Eq. (\ref{beta}) must be positive, if $c>1.5$ then $Q_c<c/(c-1)$. The distributions in Fig. \ref{fig3} generated with the pair $(c,Q_c)$ become a line when we apply a Q-logarithm function with $Q\neq Q_c$. In order to determine the value of $Q$, we compare Eqs. (\ref{rnN}) and (\ref{beta}), and therefore, we impose

\begin{equation}
\frac{c-(c-1)Q_c}{Q_c-1}=\frac{3/2-Q/2}{Q-1}.
\end{equation}

It follows

\begin{equation}
Q(c,Q_c)=\frac{(5-2c)Q_c+2c-3}{(3-2c)Q_c+2c-1},
\label{QG}
\end{equation}

hence $Q(c,1)=1$, $Q(3/2,Q_c)=Q$ and $Q(c,3)=\frac{3-c}{2-c}$.

\begin{figure}[h!] 
\centering
  \begin{subfigure}{0.49\linewidth}
  \centering
    \includegraphics[width=\linewidth]{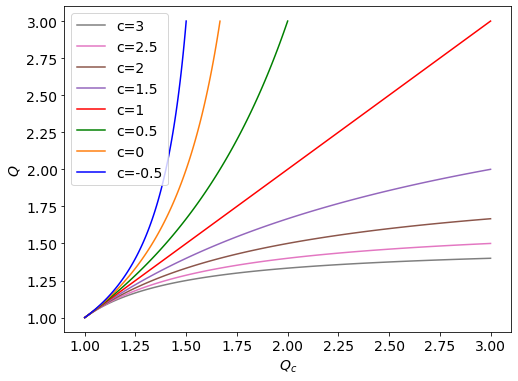}
    \caption{\label{QGvsQ}}
  \end{subfigure} 
  \hfill
    \begin{subfigure}{0.49\linewidth}
  \centering
    \includegraphics[width=\linewidth]{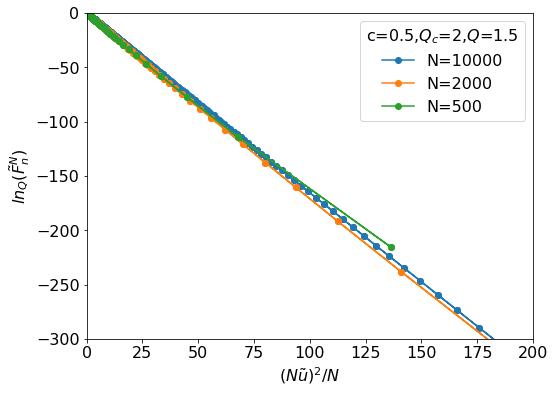}
    \caption{    \label{fig4_c05_Q25}}
  \end{subfigure}
  \caption{(a) $Q$ versus $Q_c$. Note that when $c=1.5$, $Q=Q_c$. We numerically verified for some representative cases that these values of $Q$ effectively linearize the distributions. (b) $ln_{Q}\tilde F^N_n$ versus $(N\tilde u)^2/N$. Illustration of a linearized distribution for a representative case. The factor $1/N$ in the abscissa was introduced with the aim to collapse all the curves into a single one.}
  \label{fig6}
\end{figure}

We represented Eq. (\ref{QG}) for several values of $(c,Q_c)$ in Fig. \ref{QGvsQ}, and we verified that those values of $Q$ indeed linearize the distributions, as shown in Fig. \ref{fig4_c05_Q25}.

\begin{figure}[h!] 
\centering
  \begin{subfigure}{0.49\linewidth}
  \centering
    \includegraphics[width=\linewidth]{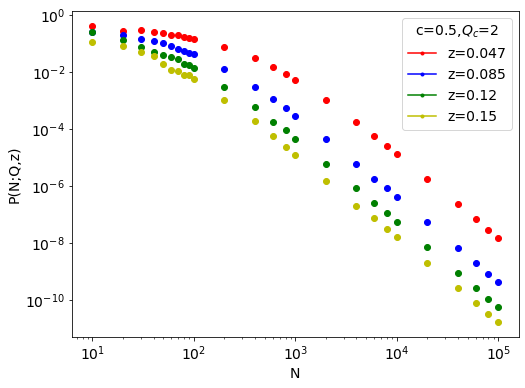}
    \caption{\label{fig5_c05_Q25} }
  \end{subfigure} 
  \hfill
  \begin{subfigure}{0.49\linewidth}
  \centering
    \includegraphics[width=\linewidth]{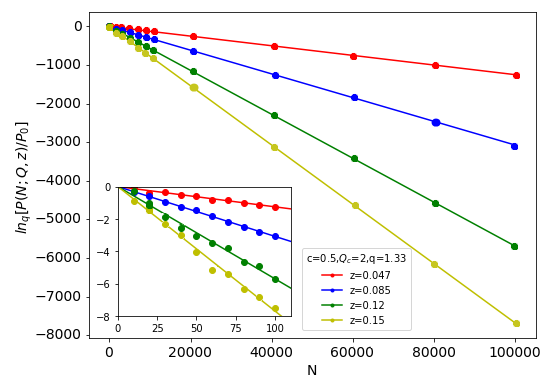}
    \caption{\label{fig6_c05_Q25} }
  \end{subfigure}
  \caption{$P(N;Q,z)$ for $c=0.5$ and $Q_c=2$ in log-log (a) and q-log (b). Note that the only distribution which provides, in all scales, straight lines in a $ln_q x$ versus $x$ representation is the q-exponential function. A linear interpolation was used to smooth the curves when values of $z$ lays between points (see Fig. \ref{fig1c05Q25}).}
  \label{fig7}
\end{figure}

 With this value of $Q$, we found that the relation $q=2-1/Q$ found in \cite{Tirnakli2022} is preserved for all values of $c$, as illustrated in Fig. (\ref{fig7})

For strongly correlated binary variables, we have

\begin{equation}
    r_q(z)=\frac{1}{q_r}\left\{\frac{1}{2}[(1+2z)^{q_r}+(1-2z)^{q_r}]-1\right\},
        \label{ex}
\end{equation}

\nd hence

\begin{equation}
    r_q(z)\sim2q_rz^2+\frac{2}{3}(3-q_r)(2-q_r)q_rz^4\;\;(z\rightarrow0).
\label{ap}
\end{equation}

\begin{figure}[h!] 
\centering
  \begin{subfigure}{0.49\linewidth}
  \centering
    \includegraphics[width=\linewidth]{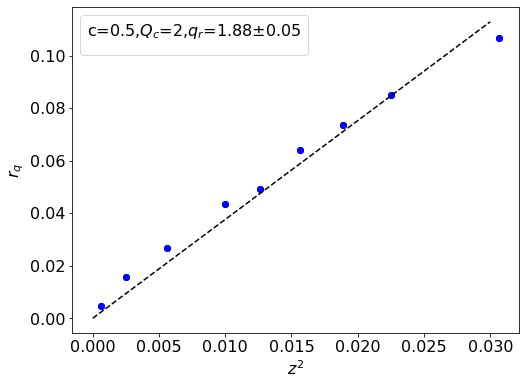}
    \caption{\label{rvsz2} }
  \end{subfigure} 
  \hfill
  \begin{subfigure}{0.49\linewidth}
  \centering
    \includegraphics[width=\linewidth]{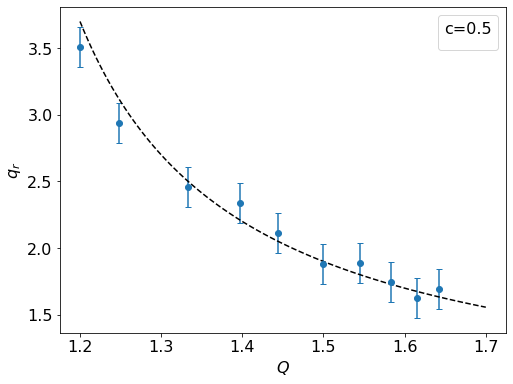}
    \caption{\label{qr} }
  \end{subfigure}
  \caption{(a) $r_q$ values calculated from Eq. \ref{qexp} are plotted as a function of $z^2$ for a representative value of $(c,Q_c)$. (b) $q_r$ versus $Q$ is plotted for the case $c=0.5$. Dots are obtained through optimization of the overall fitting of $P(N;Q,z)$ with regard to $(P_0,q_r)$ for typical values of $Q$ and various values for $z^2$. The optimization procedure uses the scipy.optimize module in Python with method 'lm'. The range of values of $Q$ plotted are related to $Q_c$ through Eq. (\ref{QG}). The dashed line represents Eq. (\ref{qreq})}
  \label{fig8}
\end{figure}

Through the optimized fitting of Fig. \ref{rvsz2}, we found the same relation as in \cite{Tirnakli2022}

\begin{equation}
q_r=\frac{7}{10}+\frac{6}{10}\frac{1}{Q-1}\;\;(1<Q<3),
\label{qreq}
\end{equation}

\nd where $Q$ is related to $Q_c$ through Eq. (\ref{QG}). This result was obtained by varying $z$ typically up to $0.175$ for different values of $Q_c$, which guarantees the verification of the dominant term in Eq. (\ref{ap}). Greater values of $z$ were not considered given our computational capacity.

The numerical determination of $P_0(Q,z)$ is much harder than that of $(q,q_r)$. A form was heuristically proposed as a simple illustration in \cite{Tirnakli2022},

\begin{equation}
    P_0(Q,z)=1/4-az^u+a(1/2-z)^u\;(0<z<1/2),
    \label{Po1}
\end{equation}

\begin{figure}[h!] 
\centering
\includegraphics[width=0.5\linewidth]{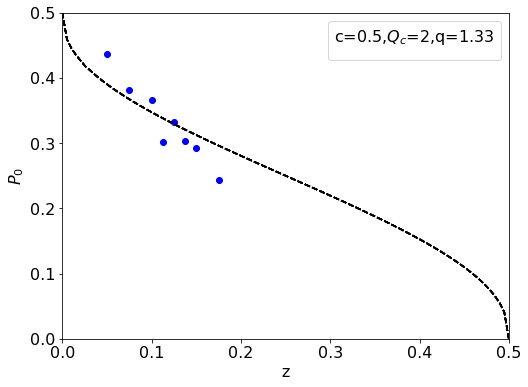}
    \caption{$P_0$ versus $z$ for an illustrative value of $Q_c$; this particular example is one of the best results we obtained for the graphic $P_0$ versus $z$. The dashed line is given by Eq. (\ref{Po1}) with the value $u=0.4$. This comparison  suggests that the heuristic expression given in \cite{Tirnakli2022} is only approximate.}
  \label{Po}
\end{figure}

\nd where $a=2^u/4$ in order to satisfy the conditions $P_0(Q, 0)=1/2$, $P_0(Q,1/2)=0$. However, Fig. \ref{Po} suggests that this analytical form is only approximate. Let us emphasize here that the exact numerical values of $P_0$ (as well as its unknown exact analytical expression) are of no central importance. In fact, they play a rather minor role in the conjecture (\ref{qexp}), much like the corresponding prefactor in the standard LDT.

The figure of the case presented here, $(c=0.5,Q_c=2)$ is one of the best figures we obtained. Other combinations of $(c,Q_c)$ produced more noisy figures. However, it appears that that the prefactor $P_0$ depends on $z$, but not on $(c,Q_c)$. 

\section{Conclusions}

To sum up, let us first recall that, in the domain of $q$-statistics based on non-additive entropies, one typically obtains a $Q$-Gaussian distribution for the velocities and a $q$-exponential weight for the energies, for $Q\geq1$, the equalities $Q=q=1$ holding precisely for the BG statistical mechanics. These generalizations should respectively reflect the corresponding generalizations of the classical Central Limit Theorem and the Large Deviation Theory. 

In the present work, we constructed a family of correlated probabilistic models and we reinforced the way towards a $q$-generalized Large Deviation Theory. We unified the two transformations $\bar q(q)$ presented in \cite{HanelThurnerTsallis2009} into a single expression, and showed that this family of models yields, in the $N\rightarrow\infty$ limit, a $Q$-Gaussian distribution for any $Q\in[1,3)$. This entire family turns out to be isomorphic to the \cite{HanelThurnerTsallis2009} model through the transformation $(c,Q_c) \to (3/2,Q)$. We presented numerical evidence that the corresponding LDT probability distribution is given by a $q$-exponential function, where $q$ is a monotonic function of $Q$ (with $Q=1$ yielding $q=1$).  
Then, we showed that this family of correlated probabilistic models reinforces the path towards a $q$-Large Deviation Theory which is consistent with the Legendre structure of thermodynamics. 

In particular, we showed the possible identification of the rate function $r_q(z)$ with a non-additive relative entropy whose index is $q_r$. A definite numerical identification of $r_q(z)$ with the $q_r$-entropy for the whole range of $z$ was not possible since our present computational capacity does not allow to increase $z$ all the way up to $z=1/2$. The $Q$-dependence $q_r(Q)$, Eq. (\ref{qreq}), as suggested in \cite{Tirnakli2022} was reconfirmed for all values of $c$. We also illustrated numerically the dependence of the prefactor $P_0$ with $z$, and its apparent independence from $(c,Q_c)$.

In the spirit of the promising results presented here, analytical approaches (or very accurate numerical approaches) would of course be very welcome, either for specific models or in the ambitious form of a $q$-generalized Large Deviation Theory based, say, on a $Q$-generalized Central Limit Theorem for an important class of strongly correlated random variables that appears frequently in physics, geophysics, astrophysics, economics, and other fields. We hope that the present numerical evidence will stimulate research on this topic and advance towards a $q$-generalized Large Deviation Theory.

\section*{Declaration of competing interest}
The authors declare that they have no known competing financial interests or personal relationships that could have appeared to influence the work reported in this paper.

\section*{Acknowledgments}
We acknowledge support from U. Tirnakli with the simulation code, as well as partial financial support from CAPES, CNPq, and FAPERJ (Brazilian agencies).

\end{document}